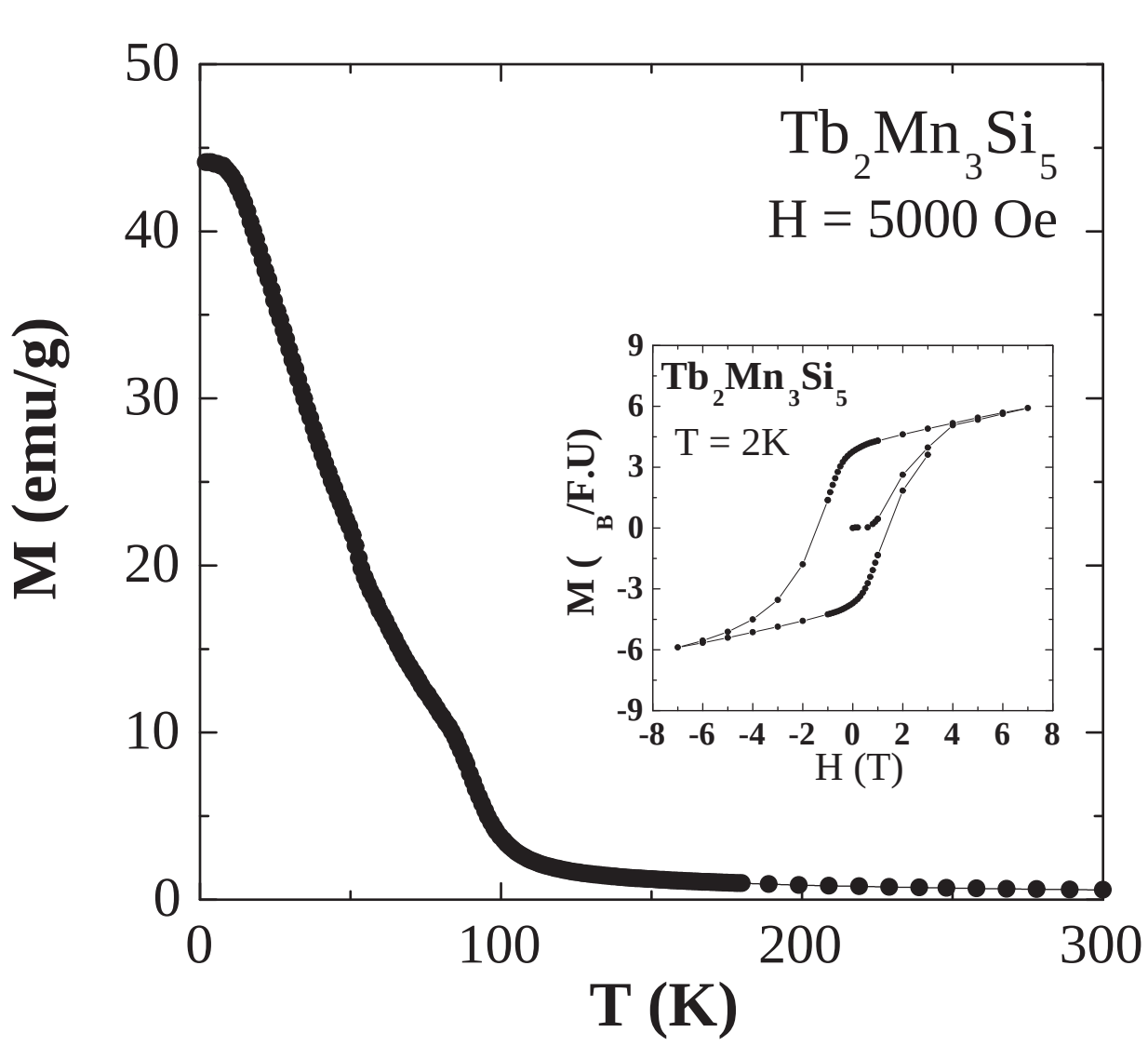

$Tb_2Mn_3Si_5$
H = 5000 Oe
M (emu/g)
T (K)
0
10
20
30
40
50
100
200
300
$Tb_2Mn_3Si_5$
T = 2K
M ( $_B$/F.U)
H (T)
9
6
3
0
-3
-6
-9
-8
-6
-4
-2
2
4
8

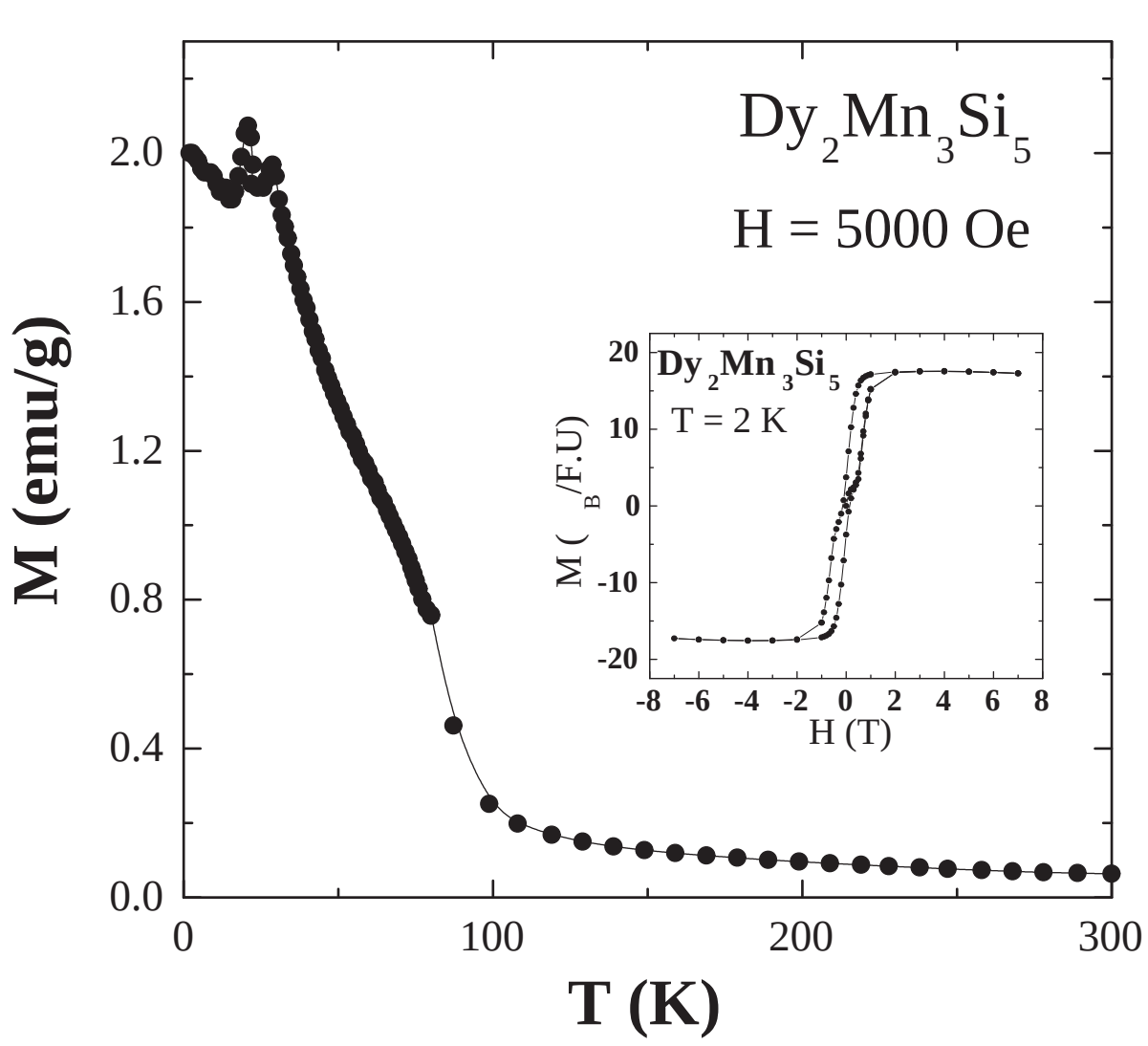

Dy2Mn3Si5
H = 5000 Oe
M (emu/g)
T (K)
0.0
0.4
0.8
1.2
1.6
2.0
0
100
200
300
Dy2Mn3Si5
T = 2 K
M (μB/F.U)
H (T)
20
10
0
-10
-20
-8
-6
-4
-2
0
2
4
6
8

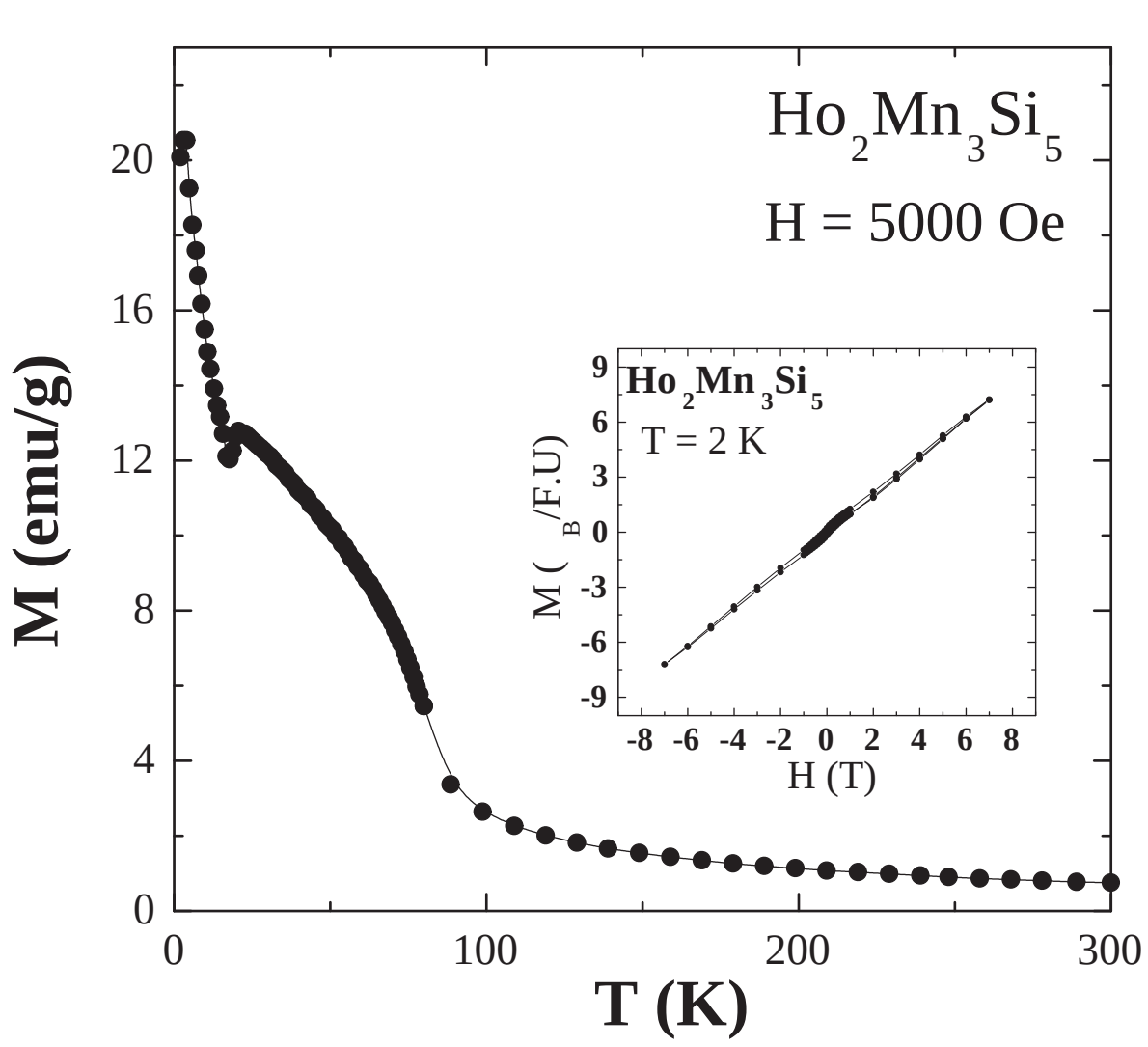

Ho2Mn3Si5
H = 5000 Oe
M (emu/g)
T (K)
Ho2Mn3Si5
T = 2 K
M ( B/F.U)
H (T)

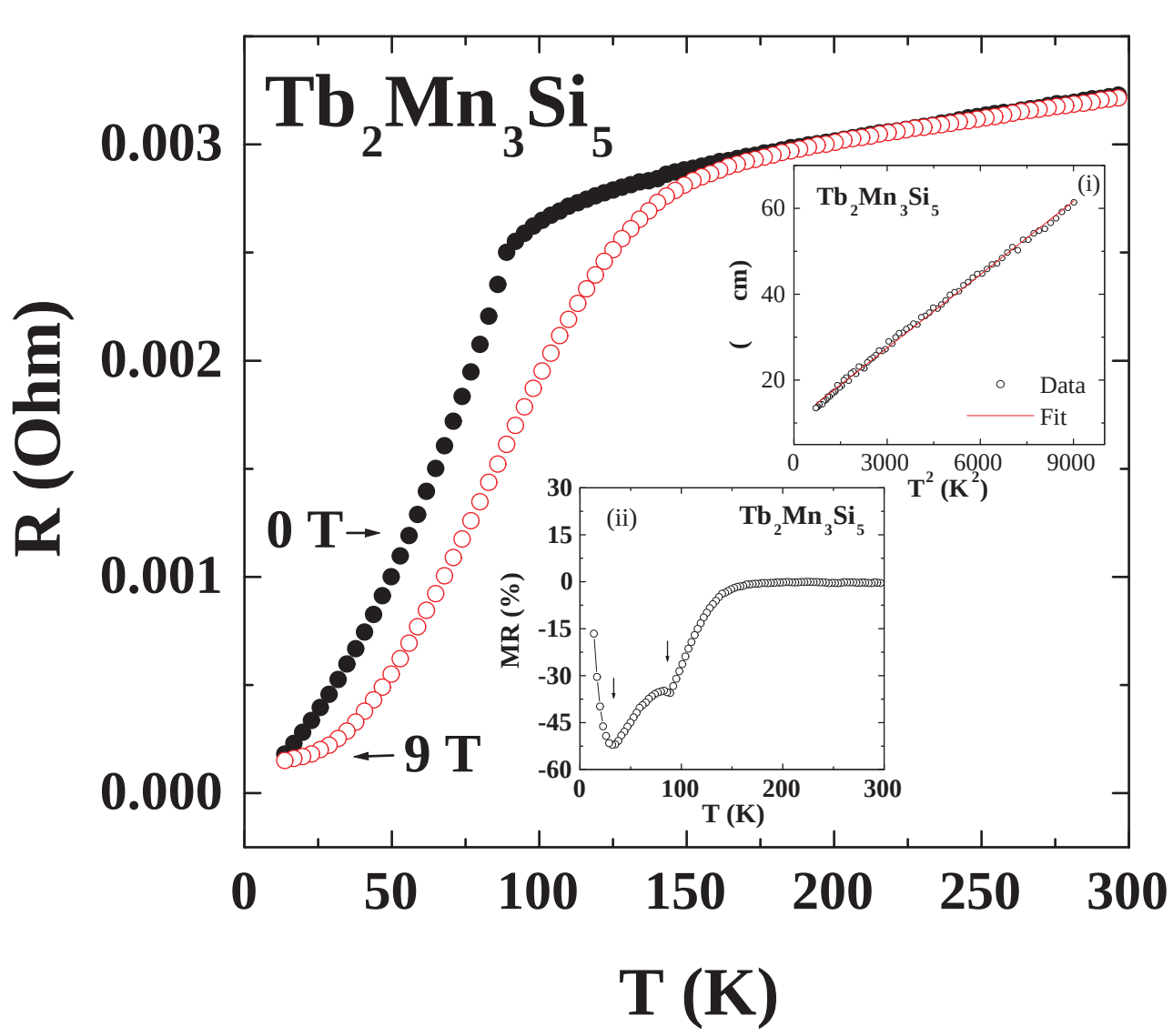

Tb2Mn3Si5
R (Ohm)
T (K)
0 T
9 T
0.000
0.001
0.002
0.003
0
50
100
150
200
250
300
(i)
Tb2Mn3Si5
Data
Fit
T2 (K2)
(ii)
MR (%)
T (K)

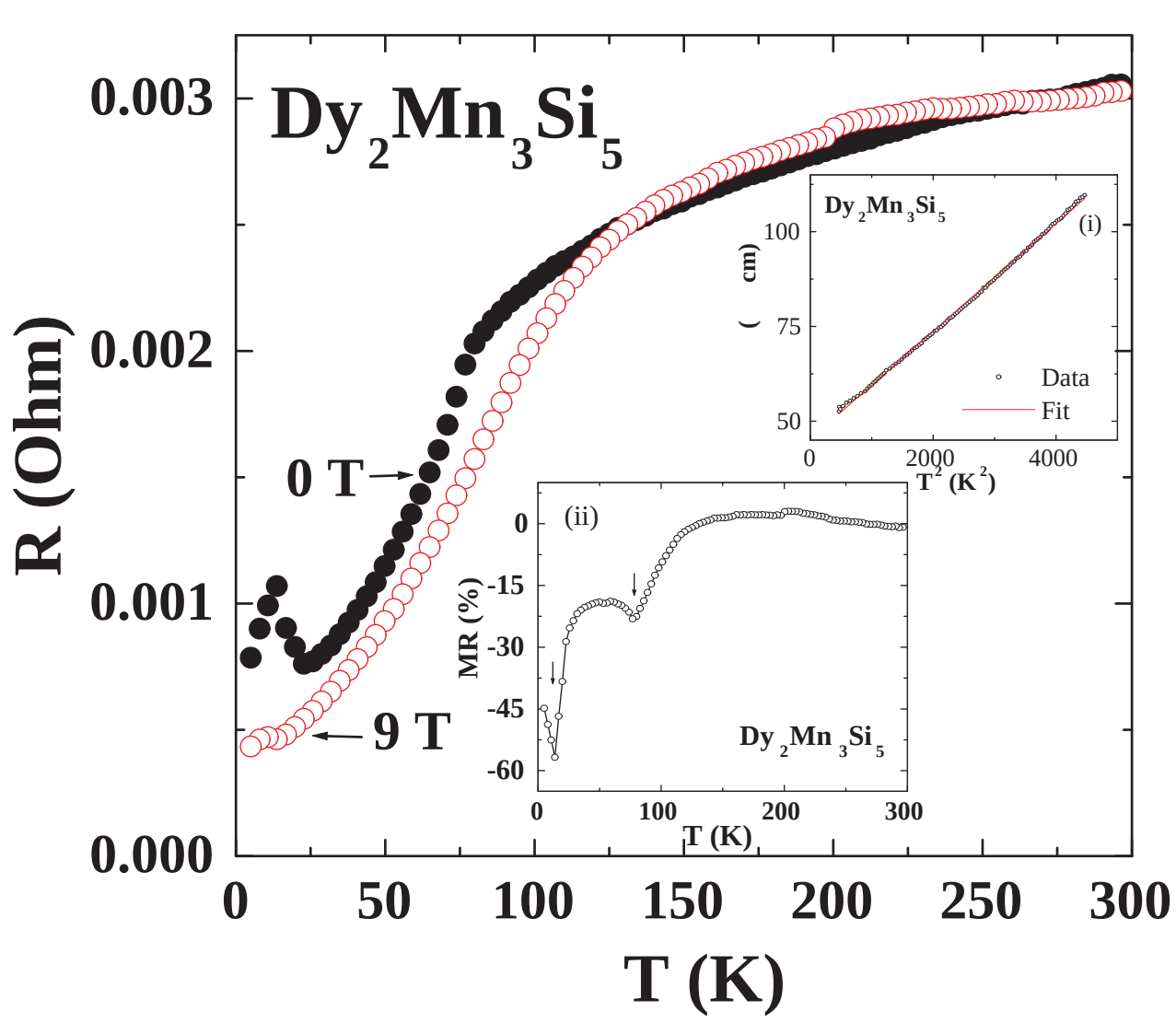

Dy2Mn3Si5
R (Ohm)
0.000
0.001
0.002
0.003
T (K)
0
50
100
150
200
250
300
0 T
9 T
(i)
(ii)
Data
Fit
MR (%)

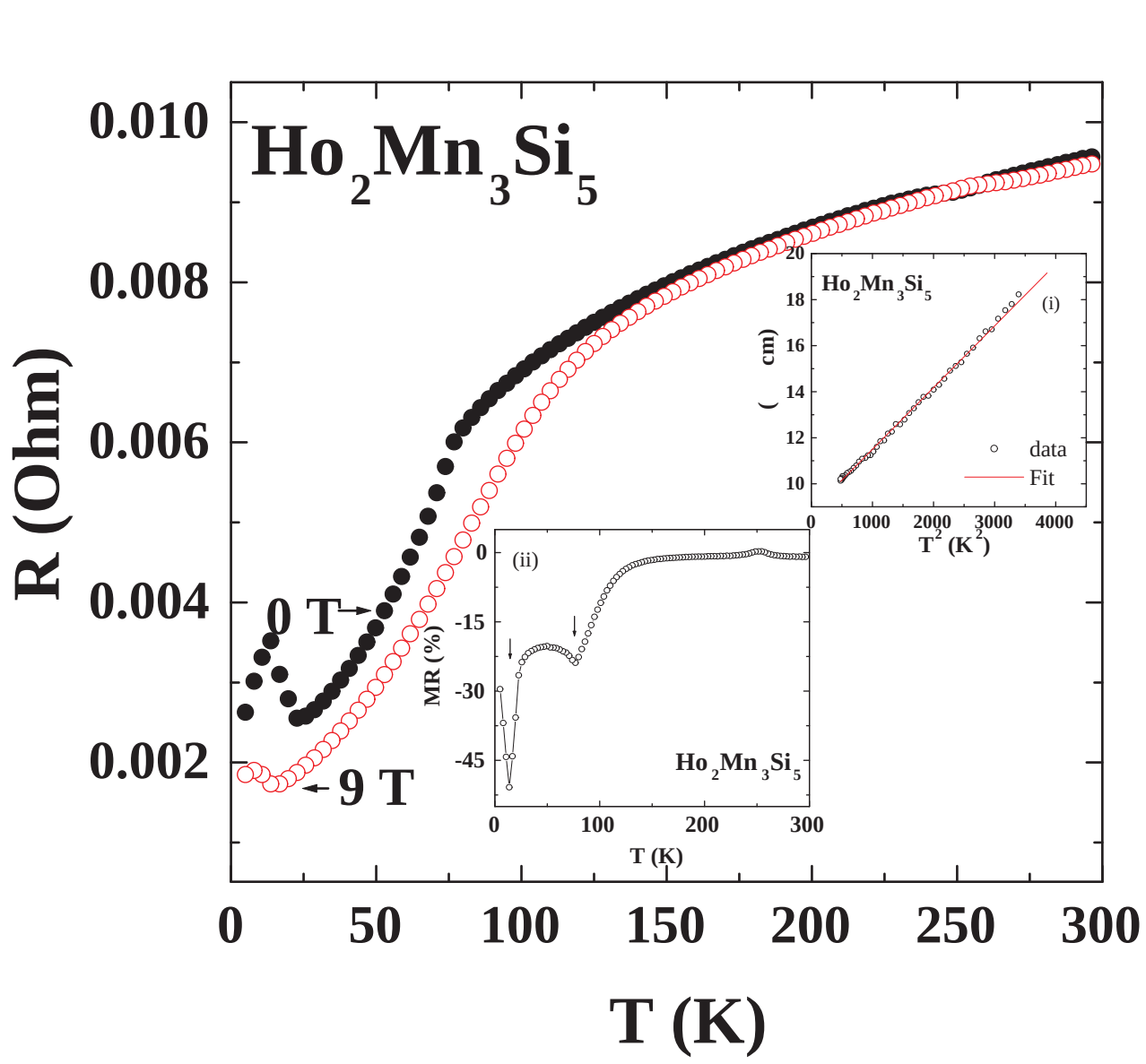

Ho2Mn3Si5
R (Ohm)
T (K)
0 T
9 T
0.010
0.008
0.006
0.004
0.002
0
50
100
150
200
250
300
(i)
data
Fit
(ii)
MR (%)

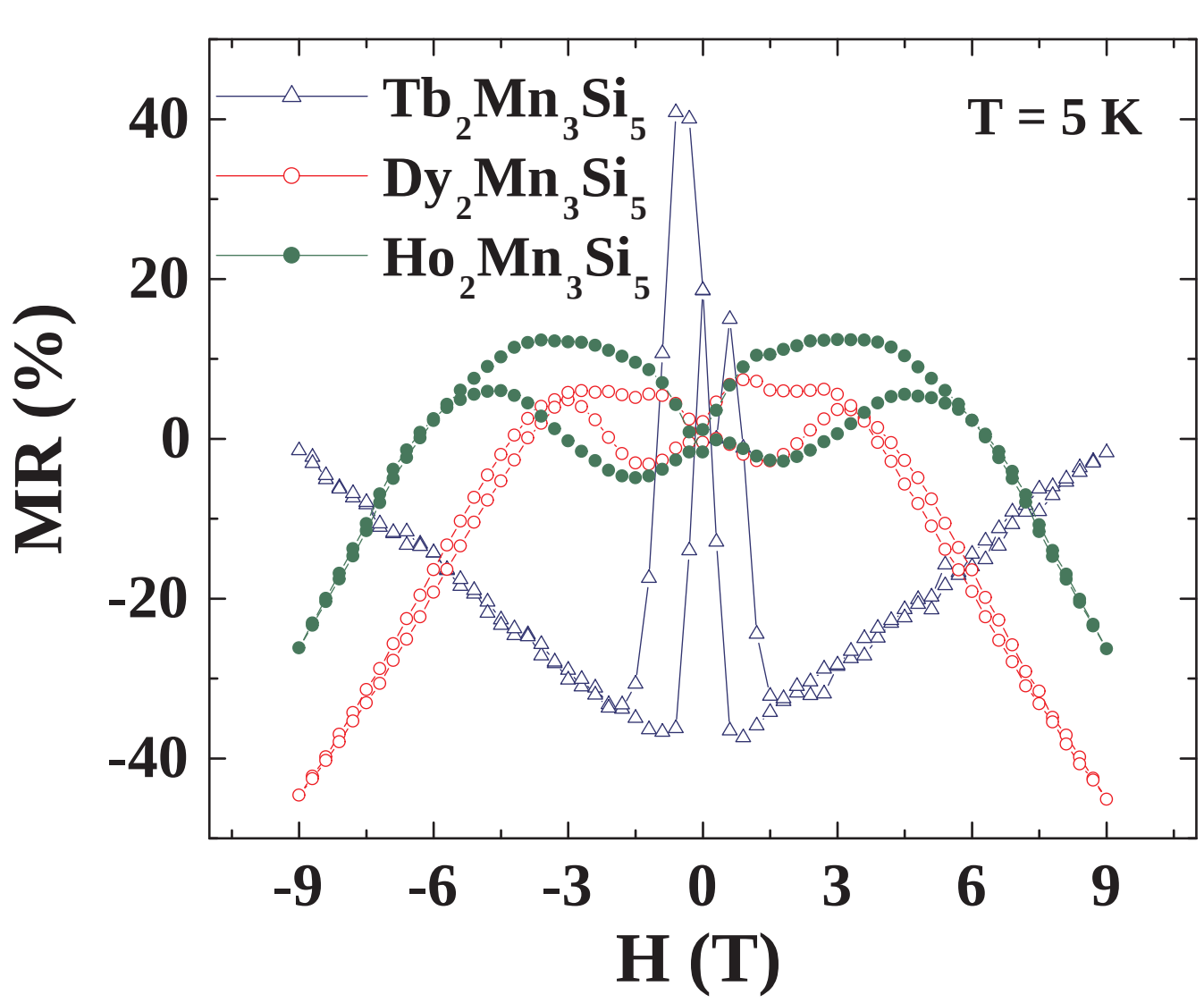

Tb2Mn3Si5
Dy2Mn3Si5
Ho2Mn3Si5
T = 5 K
MR (%)
H (T)
40
20
0
-20
-40
-9
-6
-3
0
3
6
9